\documentclass[reprint,superscriptaddress,nofootinbib,twocolumn,amsmath,amssymb,aps,pra]{revtex4-2}
\usepackage{graphicx}
\usepackage{braket}
\usepackage[colorlinks, citecolor=red, urlcolor=red, bookmarks=false, hypertexnames=true]{hyperref}
\usepackage{dcolumn}
\usepackage{bm}
\usepackage{hyperref}
\usepackage{mathrsfs}

\begin{document}
\preprint{APS/123-QED}
\title{Sensitivity Comparison of Two-photon vs Three-photon Rydberg Electrometry}
\author{Nikunjkumar Prajapati}
\affiliation{National Institute of Standards and Technology, Boulder, CO 80305, USA} 
\author{Narayan Bhusal}
\affiliation{Department of Physics, University of Colorado, Boulder, Colorado 80309, USA}
\author{Andrew P. Rotunno}

\author{Samuel Berweger}
\author{Matthew T. Simons}
\author{Alexandra B. Artusio-Glimpse}
\affiliation{National Institute of Standards and Technology, Boulder, CO 80305, USA}
\author{Ying Ju Wang}
\author{Eric Bottomley}
\author{Haoquan Fan}
\affiliation{ColdQuanta, Boulder, CO 80301, USA}
\author{Christopher L. Holloway}
\thanks{christopher.holloway@nist.gov}
\affiliation{National Institute of Standards and Technology, Boulder, CO 80305, USA}
\date{\today}

\begin{abstract}
We investigate the sensitivity of three-photon EIT in Rydberg atoms to radio frequency detection and compare it against conventional two-photon systems. Specifically, we model the 4-level and 5-level atomic system and compare how the transmission of the probe changes with different powers of the lasers used and strengths of the RF field. In this model, we also define a sensitivity metric to best relate to the operation of the current best experimental implementation based on shot noise limited detection. We find that the three-photon system boasts much narrower line widths compared to the conventional two-photon EIT. However, these narrow line features do not align with the regions of the best sensitivity. In addition to this, we calculate the expected sensitivity for the two-photon Rydberg sensor and find that the best achievable sensitivity is over an order of magnitude better than the current measured values of 5$\mathrm{\mu Vm^{-1}Hz^{-1/2}}$. However, by accounting for the additional noise sources in the experiment and the quantum efficiency of the photo-detectors, the values are in good agreement.
\end{abstract}
\maketitle

\section{Introduction}
Rydberg atoms are excited to a state with a high principal quantum number $n$ \cite{gallagher_book}. 
The large spatial separation between the electron and nucleus in this excited state produces a large polarizability and long-lived atomic state~\cite{first_ryd_elet_shaff,holl2,si_trace,gor1,Kumar2017}. 
This large polarizability makes Rydberg atoms highly sensitive to external electric (E) fields, and thus they are ideal candidates for various applications of field sensing such as metrology~\cite{7812705,si_trace,10.1117/12.2552626}, power standards~\cite{pow_stand},imaging~\cite{terehertz_imaging}, communications~\cite{8878963,doi:10.1063/1.5088821,Song:19,doi:10.1063/1.5028357,Cox_2018,9069423,8778739}, video reception~\cite{doi:10.1116/5.0098057}, and other applications \cite{Norrgard_2021,9069423, Fan_2015,gross2020ion, saffman2016quantum}. 
In addition, the Rydberg atom-based sensing is furthering directly SI-traceability via Planck's constant \cite{gor1, pow_stand, holloway2014broadband, anderson2021self,doi:10.1116/5.0097746}.

The Rydberg atom sensor typically relies on the measurement of the response of a Rydberg state to an external electric field.
This is conventionally accomplished by utilizing electromagnetically induced transparency (EIT), a coherent non-linear process that produces a narrow transparency window amid the larger Doppler absorption background at the two-photon resonance~\cite{first_ryd_elet_shaff, holloway2014broadband, holl2, si_trace, kumar2017atom}.
EIT allows for precise probing of the Rydberg state and its change in response to external fields~\cite{holl3,gallagher_book,simons2021rydberg,first_ryd_elet_shaff}. 
In the presence of an external E-field resonant with the transition between the measured Rydberg state and an adjacent Rydberg state, the EIT resonance will split proportionally to the power of the field due to the Autler-Townes effect~\cite{holl3,gallagher_book,gor1,first_ryd_elet_shaff}.

\begin{figure}[ht!]
    \centering
    \includegraphics[width=0.48\textwidth]{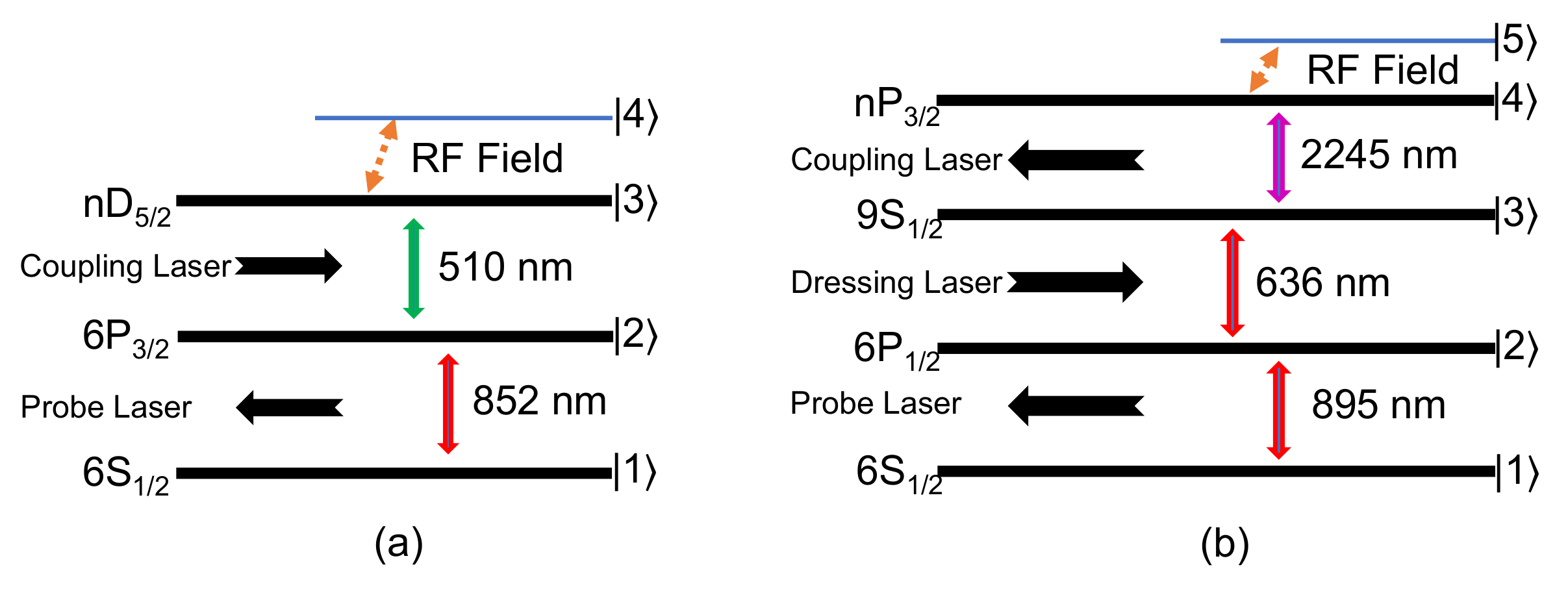}
    \caption{Rydberg atom excitation for $^{133}$Cs: (a) a typical two-photon EIT scheme, and (b) a three-photon EIT scheme. The arrows denote the relative direction of propagation in the co-linear geometry.}
    \label{figAtomLevels}
\end{figure}

Using this concept and the atomic mixer, we can measure the phase of the RF field and achieve field sensitivity down to 5$\mathrm{\mu Vm^{-1}Hz^{-1/2}}$~\cite{repump_paper,Jing2020}. 
While the E-field sensitivity can be further improved through complementary methods like population repumping~\cite{repump_paper} and meta-materials for local field enhancement~\cite{doi:10.1063/5.0088532}, sensitivity is ultimately limited in large part due to EIT linewidths.

\begin{figure*}[ht!]
    \centering
    \includegraphics[width=.8\textwidth]{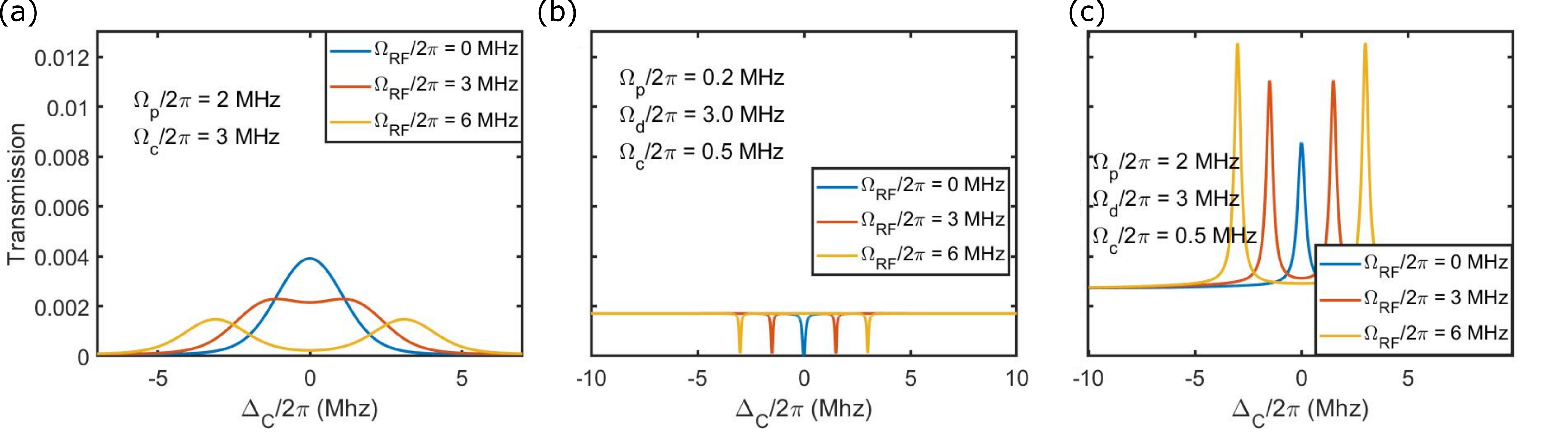}
    \caption{Comparison of two-photon and three-photon EIT spectra in Cs-133 atoms for laser Rabi rates as labeled. The different traces are for different strength radio frequency fields. (a) Two-photon EIT spectrum in Cs-133 for different applied RF Rabi rates. (b) Three-photon EIA signal resulting from using a small probe field Rabi rate. (c) Three-photon EIT signal resulting from using a large probe field Rabi rate.}
    \label{fig2phot3phot}
\end{figure*}

While the two-photon process is used to produce a substantially more narrow line compared to single-photon interactions, the EIT resonance linewidth is far from the radiative limit of the atoms. This is due to the Doppler mismatch between the probe and coupling photons used to produce the EIT, shown in Fig.~\ref{figAtomLevels} (a). The mismatch or uncancelled Doppler residual is given by~\cite{10.1117/12.2309386}
\begin{equation}
    \Gamma_{res} = \frac{\gamma_2}{|k_p|}\cdot\sum_i \vec{k}_i,
\end{equation}
where $\Gamma_{res}$ is the Doppler residual width, $\gamma_2$ is the decay rate of the first excited state $|2\rangle$, $p$ subscript is the probe optical field, and $k_i$ is the wave vector of the optical field. In Rydberg EIT, the two fields are counter propagating and co-linear to minimize this value. However, the Doppler residual is limited to roughly 3.5 MHz for the two-photon EIT in Cesium, shown by~Fig.\ref{fig2phot3phot} (b). It has been shown that by utilizing three-photon interactions, narrower linewidths can be obtained. In the case of Cesium for the D1 line (shown in Fig.~\ref{figAtomLevels} (b)), the Doppler residual can be reduced down to 39 kHz, shown by~Fig.\ref{fig2phot3phot} (b).

The implementation of the three-photon system can lead to much narrower resonances, however this does not necessarily translate directly to optimal conditions for field sensitivity, which ultimately requires an understanding of the large and complex parameter space governing this system. While other three-photon systems have been explored~\cite{PhysRevA.100.063427,10.1117/12.2309386,Carr:12}. In this manuscript, we present modeling results that identify and discuss the interplay between Rabi rates of the different fields and how they contribute to the strength of the EIT resonance, but also the switch to electromagnetically induced absorption (EIA). 
We also describe a formalism for estimating the sensitivity based on shot noise limited detection and compare the calculated sensitivity to the sensitivity limits of current two-photon experiments and determine the optimal regions of operation for the three-photon system. The description of the theoretical model is given in the appendix. 

\section{Electromagnetically Induced Absorption vs. Transmission}
Generally, spectroscopy utilizing a multi-photon process switches between resonant EIT and EIA features for even and odd numbers of optical fields, respectively \cite{mcgloin2001electromagnetically}. However, with more than two photons, the interactions become more complex. The interplay of the different Rabi rates will also play a role in whether the three-photon interaction produces an EIT or EIA resonance. Fig.~\ref{fig2phot3phot} (b) and (c) show EIA and EIT for three-photon ladder, respectively. The increase of the Rabi rate of the probe field to comparable values of the dressing field causes the EIA to switch to EIA. 

\begin{figure}
    \centering
    \includegraphics[width=1\columnwidth]{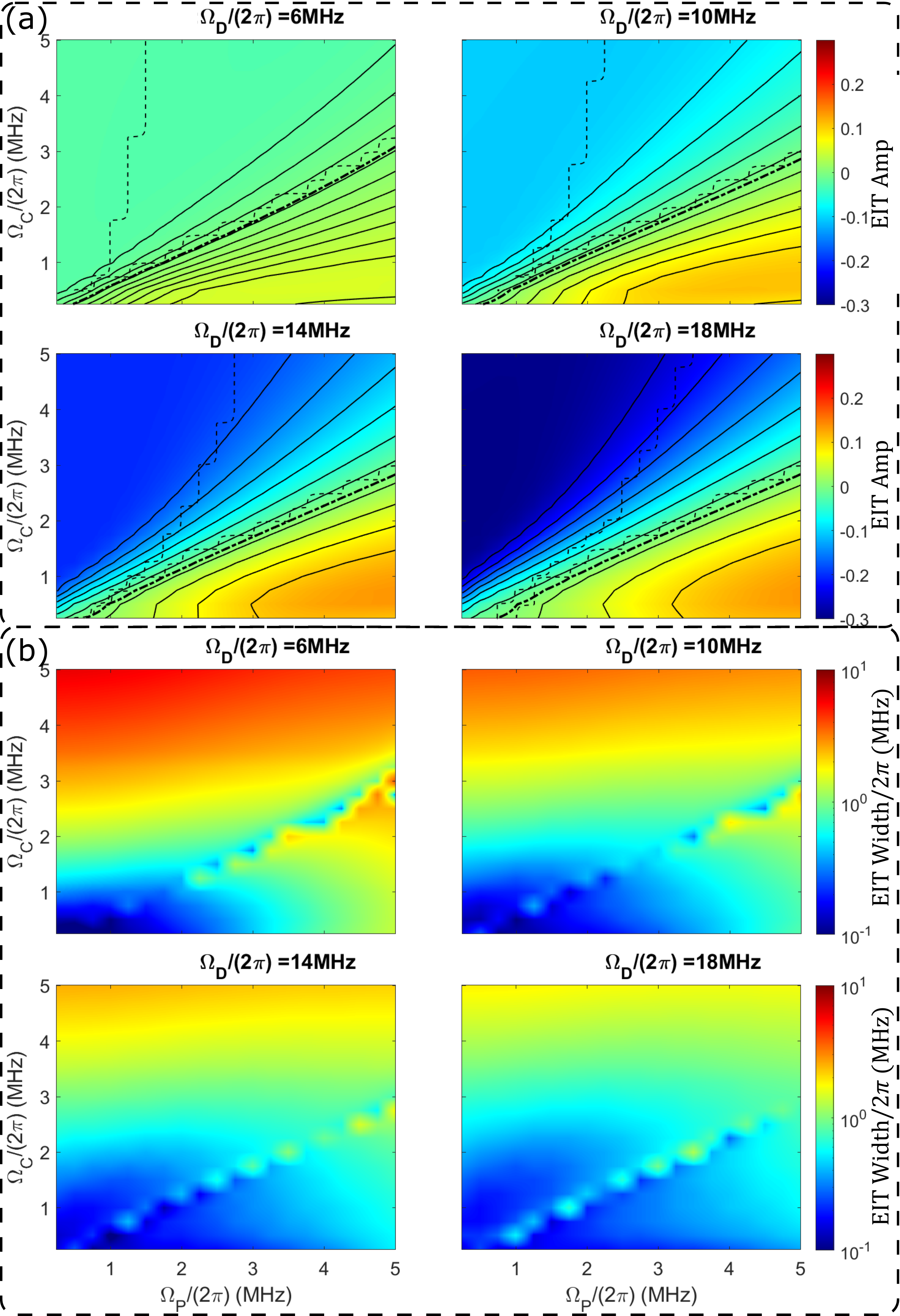}
    \caption{(a) Amplitude (A) (from Eq.~\ref{eq:amplitude}) mapped out for different probe, dressing, and coupling Rabi rates as labeled. The positive values are regions of EIT while the negative values are regions of EIA. The area within the dashed lines are regions where an applied RF field can cause a change from EIA to EIT. The bold dashed line marks when the amplitude falls to 0. (b) Full-width at half-max (FWHM) of the non-linear resonance obtained for different probe, dressing, and coupling Rabi rates as labeled.
    }\label{FigEITEIAgrid}
\end{figure}

To better understand the effects of the Rabi rates of the different optical fields, we explore the response of the nonlinear resonance for different probe, dressing, and coupling field Rabi rates, shown in Fig.~\ref{FigEITEIAgrid} (a). The EIT amplitude (A) is defined by
\begin{equation}
    A = T(\Delta_c=0)-T_{bkg},
    \label{eq:amplitude}
\end{equation}
where $T(\Delta_c=0~\mathrm{MHz})$ is the transmission value on the three-photon resonance and $T_{bkg}$ is the background transmission signal obtained the value at the beginning of the detuning trace (in other words $T(\Delta_c=-10~\mathrm{MHz})$). The cases where the EIT amplitude is negative are the regions where the signal is that of EIA rather than EIT. In Fig.~\ref{FigEITEIAgrid} (b), we show the FWHM width of the three-photon resonance for probe, dressing, and coupling Rabi rates.

We see that the Rabi rate of the dressing laser plays a large role in the general amplitude and width of the EIT. For larger dressing Rabi rates, both the EIT and EIA amplitudes increase and the crossover between the two becomes more apparent. In addition to this, the average width of the resonance begins to decrease. While the dressing Rabi rate increases the amplitude of the resonances, it does not play a large role in where the crossover from EIT/EIA occurs, defined by $\mathrm{EIT_{amp}}$ = 0 and shown by the bold dashed line in Fig.~\ref{FigEITEIAgrid} (a). This line moves very little for the different Rabi rates of the dressing field. This is likely due to the populations of the intermediate states being near zero in EIT/EIA interactions. This makes the dressing field a mediator for the process while the probe and dressing lasers largely control the relative populations of the ground and Rydberg states that determine the quality of the resonance.

\begin{figure*}[!ht]
    \centering
    \includegraphics[width=.9\textwidth]{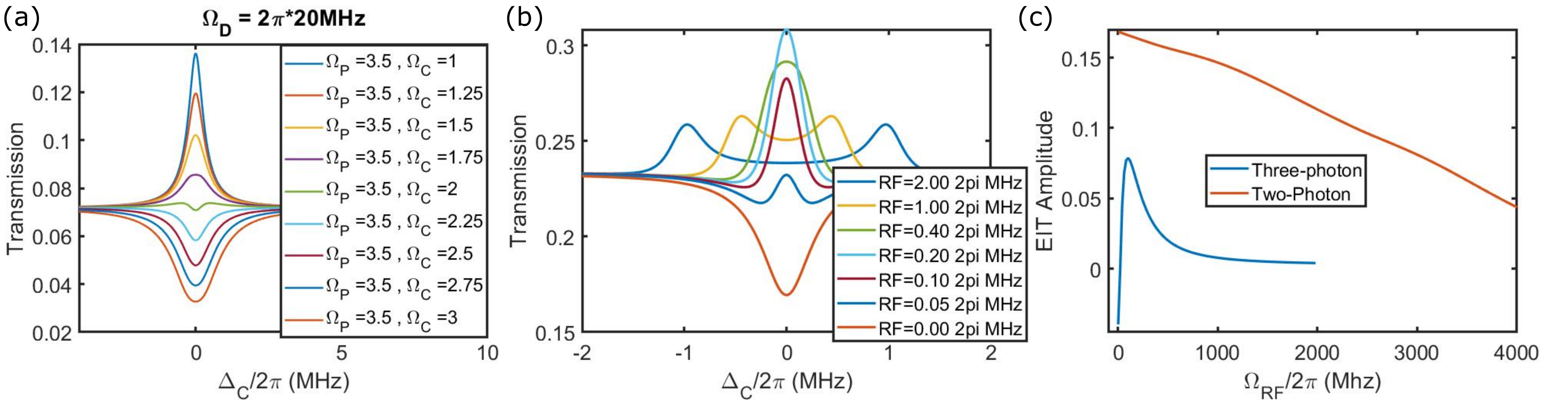}
    \caption{(a) Three-photon resonance plotted as a function of coupling laser detuning for a fixed dressing and probe laser power. The different traces represent different coupling laser Rabi rates. (b) Three-photon resonance plotted as a function of coupling laser detuning for probe, dressing, and coupling Rabi rates of 3 MHz, 20 MHz, and 2.5 MHz, respectively. The different traces are for different applied RF field strengths, as labeled in the legend. (c) Trace of resonance amplitude (A) (as defined in Eq.~\ref{eq:amplitude}) plotted against the applied RF field power for the three-photon (blue) and two-photon (red) cases. For the three-photon, the powers match those in (b). The two photon Rabi rates are 0.2 MHz probe and 12 MHz coupling.}
    \label{figEITeia}
\end{figure*}

The relative Rabi rates of the probe and coupling determine whether we sit on an EIT or EIA resonance, shown in Fig.~\ref{FigEITEIAgrid} (a). When the ratio of probe to coupling Rabi rate is small ($< 1/\sqrt(2)$), we sit on the region of EIA. This is expected for the odd order nonlinear interaction. Treating this as an effective two-photon resonance, the smaller ratio of  probe to coupling would indicate that a majority of the atoms are in the ground state. When we increase this ratio ($> 1/\sqrt(2)$), the resonance changes from EIA to EIT. A large ratio between probe to coupling typically implies the majority of the atoms are in the Rydberg state and controls the coherence of the dark state. Since the dressing laser plays a mediator role in the three-photon interaction, we look at the coherence between the ground (1) and Rydberg state (4) as a two-photon interaction~\cite{https://doi.org/10.48550/arxiv.2205.10959,2003ApPhB_76_33L},
\begin{equation}
    \rho_{14} = \frac{\Omega_p\Omega_C}{4\gamma_{2}\gamma_{4}+|\Omega_C|^2}
\end{equation}
where $\gamma_2= 2\pi\cdot5~MHz$ and $\gamma_4 = 2\pi\cdot 1.3~kHz$ are decays of the intermediate and Rydberg states, respectively. For the parameter space that we explore, the decay rates are much smaller than the coupling laser (4$\gamma_2\gamma_4<<|\Omega_C|^2$). This reduces the coherence term to $\rho_{14} = \Omega_p\Omega_C$. A strong coherence would lead to population being pumped into the dark state composed of the ground and Rydberg state. This would be synonymous with being in the region of EIT ($\Omega_p>\Omega_C$) in Fig.~\ref{FigEITEIAgrid} (a). Another interesting region is when the ratio is $\sim1/\sqrt(2)$. This is a location where the amplitude of the nonlinear resonance falls to zero, denoted by the dashed line in Fig.~\ref{FigEITEIAgrid} (a). There are regions where the linewidth seems to become substantially larger, but it is actually due to the appearance of multiple resonances that interfere and cause the fitting to fail, shown in Fig.~\ref{FigEITEIAgrid} (b) and shown in Fig.~\ref{figEITeia} (a).

Near the locations of the crossovers, the response to radio frequency (RF) fields also changes considerably. Introducing an RF field in Rydberg atoms typically results in Autler-Townes (AT) splitting that scales with the strength of the field. However, in the three-photon configuration near the transition region, the applied RF field can actually cause a change from EIA to EIT and then begin to split, shown by Fig.~\ref{figEITeia} (b) and represented by the region in the dashed lines in Fig.~\ref{FigEITEIAgrid} (a). This region is likely due to the splitting of the Rydberg state reducing the effective Rabi rate of the coupling laser. This will push the coherence conditions such that the effective coupling Rabi rate is smaller than the probe, causing the transition from EIA to EIT by the RF. This region of RF induced flipping seems to offer a large change in the transmission amplitude for small changes in the RF field or in other words has a larger slope which also corresponds to sensitivity, shown in Fig.~\ref{figEITeia} (c). This figure shows how the amplitude (A) changes with an applied RF field for the two (red) and three (black) photon cases. One can immediately see the difference in the slope between the two-photon and three-photon cases. But even more than that, there is a large difference in slope when moving through the EIT/EIA transition compared to when the peaks begin to split. To compare the different regions of the three-photon system and identify the ideal operating conditions, in the next section we present an estimation of the sensitivity of the system based on shot noise limited detection and observation of a weak signal on top of a local oscillator (LO). This is how the experiments are performed and should provide a good estimation of the expected sensitivities.

\section{Sensitivity Estimation}
In experimental implementations, the best sensitivity has been achieved by utilizing the atom-based heterodyne. This measurement relies on using an LO RF field to split the resonance such that the response to an additional field will be maximized, in other words the location of the maximum slope in Fig.~\ref{figEITeia} (c). Note the large difference in response between the two-photon and three photon cases. However, even more notable is the response of the three photon system when transitioning from EIA to EIT. The sensitivity then boils down to how small of a change in transmission in response to an RF we can measure. This is governed by the noise limitations of detection. In recent work, it has been found that the fundamental limit for the detection is the shot noise limit. This is what we use as the limit here. We begin by defining the laser power after the atoms in terms of the transmission,
\begin{equation}
    P_{out} = P_0\cdot T
\end{equation}
where $T$ is the transmission and $P_o$ is the probe power. When performing minimum field strength measurements, we are looking at how the output probe power changes with small changes in the RF power. As these RF changes (modulation) becomes smaller, they falls into the noise. These changes are similar to how a lock-in amplifier would extract a signal from the atoms acting as a mixer. The shot noise is given by,
\begin{equation}
    \sigma = \sqrt{n}
    = \sqrt{\frac{P_0\cdot \tau}{hf}},
\end{equation}
where $h$ is Planks constant, $\tau$ is the collection time, and $f$ is the laser frequency. The noise power of the laser is then found by multiplying the number of noise photons by the energy per photon,
\begin{equation}
    P_{noise} = \sigma\cdot \frac{hf}{\tau} = \sqrt{\frac{P_0 hf}{\tau}}
\end{equation}
The minimum resolvable change in transmission occurs when the signal ($\Delta P_0$) is equal to the noise power,
    \begin{align*}
        \Delta P_{out} &= P_{noise},\\
        P_0 \Delta T_{min} &= \sqrt{\frac{P_0 hf}{\tau}}, \\
        \Delta T_{min} &= \sqrt{\frac{hf}{P_0\tau}}
    \end{align*}
where $\Delta T_{min}$ is the minimum resolvable transmission change for a given probe field power. We can easily convert this to be in terms of the probe Rabi frequency
\begin{equation}
    P_0 = \frac{c\epsilon_0\Omega_p^2 h^2 A}{2\wp_{12}^2},
\end{equation}
where $\wp_{12}$ is the transition dipole moment of the probe transition, $c$ is the speed of light, $h$ is planks constant, $\epsilon_0$ is the free space permittivity, f is the frequency of the probe light, $A$ is the beam area (1 mm beam radius), and $\Omega_p$ is the probe Rabi frequency$/2\pi$.

Finally, the minimum resolvable transmission is,
\begin{equation}
    \Delta T_{min} = \frac{\wp_{12}\sqrt{f\cdot2}}{r\Omega_p\sqrt{h\cdot\epsilon_0\cdot c\cdot\pi\cdot \tau}},
    \label{eq:delTmin}
\end{equation}

We then multiply the minimum resolvable transmission by the inverse of the maximum slope of the transmission for changing the Rabi strength of the RF field. This will give us the smallest measurable Rabi rate for the RF field.
\begin{equation}
    \Omega_{RF} = \Delta T_{min}\cdot\left(\frac{\Delta\Omega_{RF}}{\Delta T}\right)_{max}
    \label{eq:sensitivity}
\end{equation}

This can be used to scale the state dependant sensitivity to E fields using,
\begin{equation}
    E_{RF} = \frac{\Omega_{RF} \cdot h}{\wp_{RF}},
    \label{eq:rf_to_rabi}
\end{equation}
where $\wp_{RF}$ is the transition dipole moment between nearby Rydberg states.

\begin{figure}
    \centering
    \includegraphics[width=1\columnwidth]{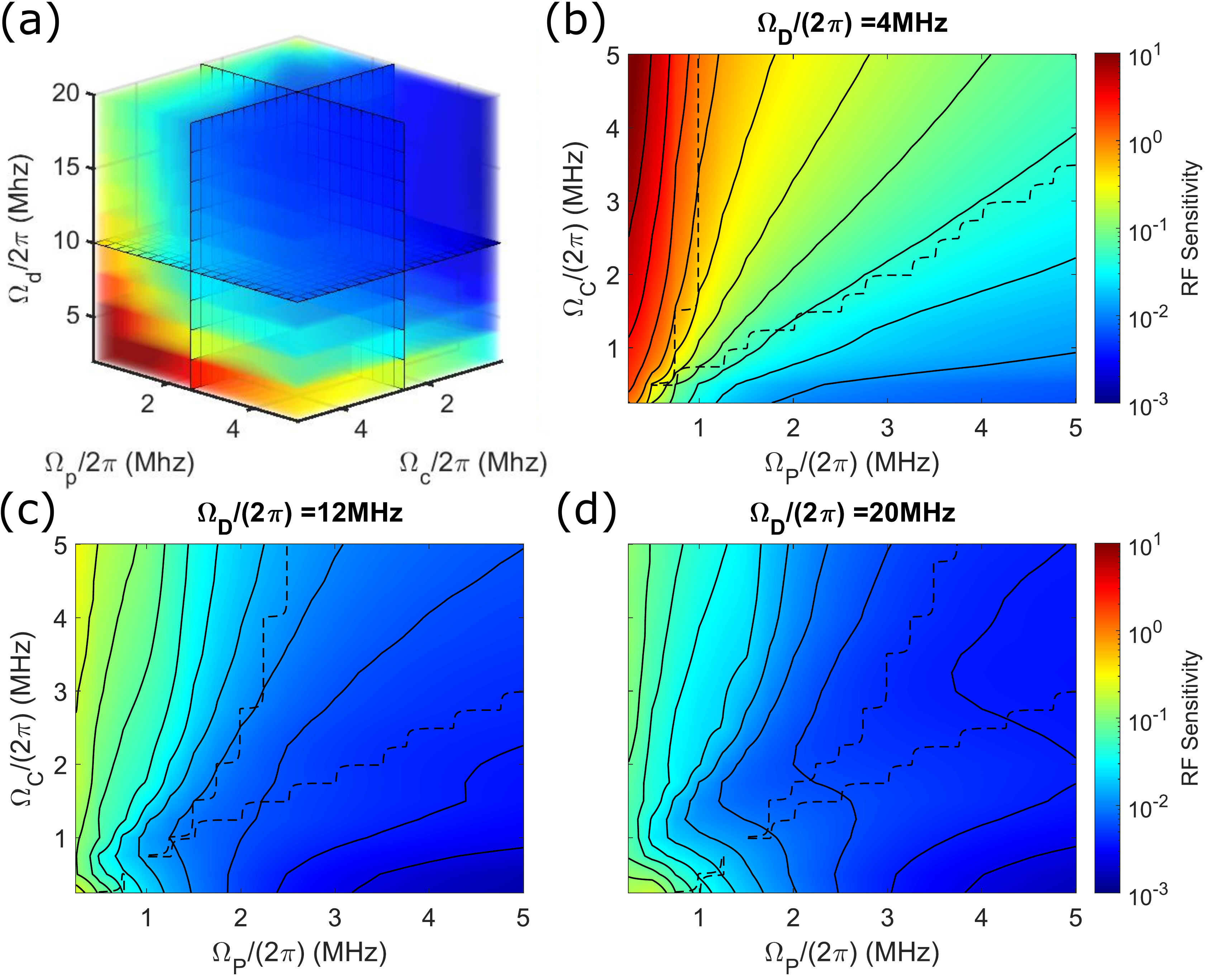}
    \caption{RF sensitivities (given by Eq.~\ref{eq:sensitivity}) calculated at various combinations of optical field strengths. The Rabi frequency for the dressing field is varied from $2\pi\times4 \text{ MHz}$ to $2\pi\times20 \text{ MHz}$ in the interval of $2\pi\times2 \text{ MHz}$. (a) Sensitivity in 3-D figure that shows dependence on probe, dressing, and coupling Rabi rates. (b-d) Sensitivity slices from (a) plotted against the probe and coupling Rabi rates for different dressing Rabi rates, as labeled. The area within the dashed lines are regions where an applied RF field causes a change from EIA to EIT.}
    \label{FigSenGrid}
\end{figure}

Using this definition for sensitivity, we can now apply it to the model. For every combination of Rabi rates shown in Fig.~\ref{FigEITEIAgrid} (a), we model the RF response of the EIT amplitude and extract the maximum slope. For the three-photon and two-photon cases, the EIT amplitude change with RF Rabi Rate is shown by Fig.~\ref{figEITeia} (c). We then use Eq.~\ref{eq:delTmin} and Eq.~\ref{eq:sensitivity} to determine the sensitivity and map out the sensitivity for different probe, dressing, and coupling Rabi rates.

\section{Two-photon vs. Three-photon RF Sensitivity}

The sensitivity is calculated for the two-photon and three-photon systems. We identify the optimal regions of operation for the three-photon system. We then compare the calculated sensitivity of the two-photon system to current best-effort experimental results and to the calculated sensitivity of the three-photon system. We also present potential problems with the characterization of the optimal region for the three-photon system.

\begin{figure}[!ht]
   \centering
    \includegraphics[width = 1\columnwidth]{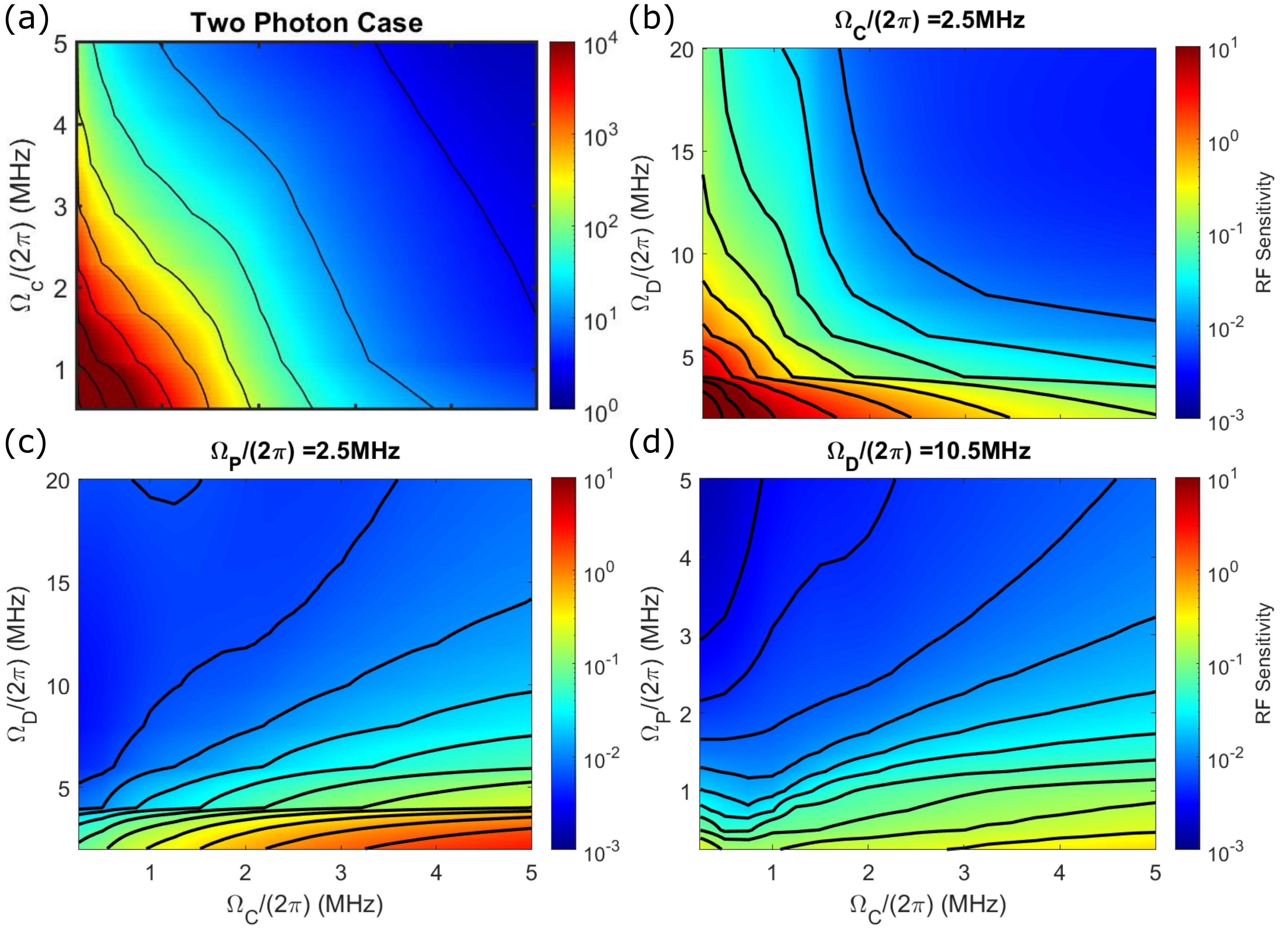}
    \caption{(a) Calculated RF sensitivities for two-photon excitation as a function of probe and coupling Rabi rate. Calculated RF sensitivities for three photon excitation for (b) fixed coupling $\Omega_P=2\pi\times$2.5~MHz while sweeping through dressing and probe Rabi rates, (c) fixed probe $\Omega_P=2\pi\times$2.5~MHz while sweeping through dressing and coupling, and (d) fixed dressing $\Omega_P=2\pi\times$10.5~MHz while sweeping through probe and coupling.}
    \label{figRFsens}
\end{figure}

The RF sensitivity dependence on the Rabi rates of the probe, dressing, and coupling lasers is shown in Fig.~\ref{FigSenGrid}. We find that the sensitivity generally improves for increased Rabi rate of the probe and dressing fields. However, larger coupling fields reduce the sensitivity of the sensor to RF fields. This is likely due to the fact that the width of three-photon resonance increases substantially with coupling Rabi rate compared to the probe Rabi rate, while the amplitude of the resonance increases similarly, shown in Fig~\ref{FigEITEIAgrid} (a) and (b). It should be noted that the crossover region, within the dashed lines in Fig.~\ref{figEITeia} (a) and Fig.~\ref{FigSenGrid}, are not indicative of the most optimal region of operation. The crossover induced by the RF field shows a large slope, but the amplitude of the resonance is generally small in those regions. This makes it non-ideal for sensitive measurements. The most sensitive region (i.e. smaller values) occurs for small coupling laser powers, large probe laser powers, and larger dressing powers, shown in Fig.~\ref{figRFsens}.

Another important factor that we examined was the line width dependence on and how it relates to sensitivity. While the assumption was that the narrower line region provides a better sensitivity, this was not entirely the case. The line width of the resonance was an important factor, but the ideal sensitivity occurred when the amplitude was maximized while maintaining a small line width. The best sensitivity was seen when the linewidth was almost 1 MHz.

In the two-photon case, the best sensitivity reported to date was 5$\mathrm{\mu Vm^{-1}Hz^{-1/2}}$~\cite{repump_paper}. This was done by using the atom based mixer and was limited in detection by atomic noise ($\sim$10 dB above shot noise) that was imparted onto the probe beam. We can convert this sensitivity into an effective Rabi rate using Eq.~\ref{eq:rf_to_rabi}. This gives us a sensitivity of $\sim$100 Hz$/\sqrt{Hz}$ in Rabi rate. The experiment was performed for a probe Rabi rate of 5 MHz and a coupling Rabi rate of 0.5 MHz. The sensitivity calculations for the same probe and coupling strengths yield a sensitivity of a Rabi rate of 7 Hz$/\sqrt{Hz}$, shown in Fig.~\ref{figRFsens} (a). The expected sensitivity from the calculation is over 10 times smaller than the measured value. However, this agrees with the experimental limitations since this calculation relies on shot noise limited detection and 100\% quantum efficient photodiodes. The experiment was performed using 60\% quantum efficient diodes and a system noise that was 10 dB above the shot noise even with differential detection. This makes the sensitivity calculation for the two-photon system in good agreement with the experimental system.

The difference in sensitivity between the two-photon system and the three-photon system is substantial. Fig.~\ref{figRFsens} (b)-(d) show the RF sensitivity in Rabi rate Hz$/\sqrt{Hz}$ of the three-photon system for three different cases: b) fixed coupling laser power for different probe and dressing laser powers, c) fixed probe laser power for different coupling and dressing laser powers, and d) fixed dressing laser for different probe and coupling laser powers. The best calculated sensitivity of the three-photon system is on the order of $10^-3$ Hz$/\sqrt{Hz}$. For the two-photon system, the best calculated sensitivity is roughly 1 Hz$/\sqrt{Hz}$. This corresponds to three orders of magnitude improvement between the two systems. However, there are certain limitations of what is possible to achieve for the lasers. The Rabi rate is limited in the coupling laser in the current two-photon systems, the Rabi rates of the 636 nm laser and 2 $\mathrm{\mu m}$ lasers are also limited by the small dipole matrix elements. For the optimal case in the three-photon system, the Rabi rate of the probe, dressing, and coupling lasers are 5 MHz, 12 MHz, and 0.5 MHz, respectively. For a beam width of 1 mm, this corresponds to laser powers of 20~$\mathrm{\mu W}$ for the probe, 40 mW for the dressing, and 10 mW for the coupling laser. In the case of the three-photon system, there are limitations to power of the 636 nm laser.

While we have identified an ideal region of operation, this is done without considering several aspects that could play a role in the sensitivity. This includes the key assumption of shot noise limited detection and the fact that the shot noise is purely from the probing laser and does not include noise from the coupling or dressing lasers. While it has been demonstrated that the noise from the coupling laser does not add to the probe laser noise in the two-photon configuration~\cite{repump_paper}, it has yet to be explored experimentally for the three-photon case. Since the dressing laser and coupling lasers in the three photon configuration have much larger dipole moments than in the two-photon case, there is more of a chance for the noise to be transferred from the dressing laser onto the probe. In addition to this, we did not take into account the effects of transit time broadening. For a beam waist of 1 mm that we assumed, we would expect a transit time of roughly 6 $\mathrm{\mu s}$. This is rather fast and will set a bound to the minimum line width in the system to be roughly 100 kHz. However, since the optimal sensitivity does not occur at the smallest line width, this should not be a large effect.

\section{Conclusion}
Electromagnetically induced transparency in Rydberg atoms is an important technique to measure RF fields over a wide frequency range. We investigated a three-photon excitation method and showed that this method substantially reduces the effects of residual Doppler broadening in comparison to the conventional two-photon approach. Our theoretical investigation shows that the three-photon EIT in $^{133}$Cs can be utilized to measure RF fields more precisely due to the narrowing of EIT peaks. In addition, the three-photon EIT responds very quickly to the change in applied RF field strength, improving our measurement sensitivity. In other words, the three-photon excitation method allows us to access/measure very low-power RF fields. More precisely, we have shown that the reduced Doppler broadening in the three-photon scheme in $^{133}$Cs translates to improved RF sensitivity of almost an order magnitude. We also observed interesting EIT/EIA transition phenomena in which the transmission peak moves swiftly to the other side of the baseline as we slowly increase the RF power, flipping the EIT signal to EIA and vice versa. This feature makes our three-photon atomic sensor more responsive to weak RF fields. We demonstrated that we can utilize this exaggerated atomic response to enhance the measurement sensitivities by almost an order magnitude. These improvements in RF sensitivity and resolution have huge implications on our ability to implement RF field-based applications such as communication, metrology, and imaging.

\section{Appendix: Theoretical Model}

The benefit of having a theoretical model is that it allows us to assess a large parameter space in order to investigate how these parameters affect the sensitivity and the overall performance of these Rydberg sensor systems. We start by noting that the power of the probe beam measured on the detector (i.e., the probe transmission through the vapor cell) is given by \cite{berman2011book}
\begin{equation}
P=P_0 \exp\left(-\frac{2\pi L \,\,{\rm Im}\left[\chi\right]}{\lambda_P}\right)=P_0 \exp\left(-\alpha L\right) \,\,\, ,
\label{intensity}
\end{equation}
where $P_0$ is the power of the probe beam at the input of the cell, $L$ is the length of the cell, $\lambda_P$ is the wavelength of the probe laser,  $\chi$ is the susceptibility of the medium seen by the probe laser, and $\alpha=2\pi{\rm Im}\left[\chi\right]/\lambda_P$ is Beer's absorption coefficient for the probe laser.  The susceptibility for the probe laser is related to the density matrix component ($\rho_{21}$)  by the following \cite{berman2011book}
\begin{equation}
\chi=\frac{2\,{\cal{N}}_0\wp_{12}}{E_P\epsilon_0} \rho_{21_D} =\frac{2\,{\cal{N}}_0}{\epsilon_0\hbar}\frac{(d\, e\, a_0)^2}{\Omega_P} \rho_{21_D}\,\,\, ,
\label{chi1}
\end{equation}
where $d=2.02$\cite{SteckCsData} is the normalized transition-dipole moment for the probe laser, $\Omega_P$ is the Rabi frequency for the probe laser in units of rad/s. The subscript $D$ on $\rho_{21}$ presents a Doppler averaged value. ${\cal{N}}_0$ is the total density of atoms in the cell and is given by
\begin{equation}
{\cal{N}}_0= \frac{p}{k_B Temp} \,\, ,
\label{nn}
\end{equation}
where $k_B$ is the Boltzmann constant, $Temp$ is temperature in Kelvin, and the pressure $p$ (in units of Pa) is given by \cite{SteckCsData}
\begin{equation}
p=10^{9.717-\frac{3999}{T}} 
\label{ppp}
\end{equation}
In eq. (\ref{chi1}), $\wp_{12}$ is the transition-dipole moment for the $\ket{1}$-$\ket{2}$ transition, $\epsilon_0$ is the vacuum permittivity, and $E_P$ is the amplitude of the probe laser E-field.

The density matrix component ($\rho_{21}$) is obtained from the master equation \cite{berman2011book}
\begin{equation}
\dot{\rho}=\frac{\partial \rho}{\partial t}=-\frac{i}{\hbar}\left[H,\rho\right]+{\cal{L}} \,\,\, ,
\label{me}
\end{equation}
where $H$ is the Hamiltonian of the atomic system under consideration and ${\cal{L}}$ is the Lindblad operator that accounts for the decay processes in the atom.

For the five-level system shown in Fig.~\ref{figAtomLevels}, the Hamiltonian can be expressed as:
\begin{equation}
\begin{footnotesize}
H=\frac{\hbar}{2}\left[\begin{array}{ccccc}
0 & \Omega_P & 0 & 0&0\\
\Omega_P & A & \Omega_D & 0&0\\
0 & \Omega_D & B & \Omega_{c}&0\\
0 & 0 & \Omega_{c} & C &\Omega_{RF}\\
0 & 0 &0& \Omega_{RF} & D\\
\end{array}
\right]\,\, ,
\end{footnotesize}
\label{H4}
\end{equation}
where $\Omega_P$, $\Omega_D$, $\Omega_{c}$, $\Omega_{RF}$ are the Rabi frequencies of the probe laser, dressing laser, coupling laser, and RF field, respectively. Also,
\begin{equation}
\begin{array}{rcl}
A&=&2\Delta_P \\
B&=&-2(\Delta_P+\Delta_D)\\
C&=&-2(\Delta_P+\Delta_D+\Delta_C)\\
D&=&-2(\Delta_P+\Delta_D+\Delta_C+\Delta_{RF}) ,\\
\end{array}
\end{equation}
where $\Delta_P$, $\Delta_D$, $\Delta_D$, and $\Delta_{RF}$ are the detunings of the probe laser, dressing laser, couple laser, and the RF field, respectively, defined as
\begin{equation}
\Delta_{p,d,c,RF}=\omega_{p,d,c,RF}-\omega_{12,23,34,45} \,\,\, ,
\label{detuningeq}
\end{equation}
where $\omega_{12,23,34,45}$ are the on-resonance angular frequencies for the probe, dressing, coupling, and RF fields, respectively.

\vspace{-3mm}
\begin{equation}
\resizebox{0.48\textwidth}{!}{$
\mathcal{L}=\left[\begin{array}{ccccc}
\Gamma_{2} \rho_{22} & -\gamma_{12} \rho_{12} & -\gamma_{13} \rho_{13} & -\gamma_{14} \rho_{14} & -\gamma_{15} \rho_{15} \\
-\gamma_{21} \rho_{21} & \Gamma_{3} \rho_{33}-\Gamma_{2} \rho_{22} & -\gamma_{23} \rho_{23} & -\gamma_{24} \rho_{24} & -\gamma_{25} \rho_{25} \\
-\gamma_{31} \rho_{31} & -\gamma_{32} \rho_{32} & -\Gamma_{3} \rho_{33} & -\gamma_{34} \rho_{34} & -\gamma_{35} \rho_{35} \\
-\gamma_{41} \rho_{41} & -\gamma_{42} \rho_{42} & -\gamma_{43} \rho_{43} & \Gamma_{3} \rho_{33}-\Gamma_{4} \rho_{44} & -\gamma_{45} \rho_{45} \\
-\gamma_{51} \rho_{51} & -\gamma_{52} \rho_{52} & -\gamma_{53} \rho_{53} & -\gamma_{45} \rho_{45} & \Gamma_{4} \rho_{44}-\Gamma_{5} \rho_{55}
\end{array}\right]$}
\label{L4}
\end{equation}

For the five-level system, the ${\cal{L}}$ matrix is given in eq.~(\ref{L4}), where $\gamma_{ij}=(\Gamma_i+\Gamma_j)/2$ and $\Gamma_{i, j}$ are the transition decay rates. Since the purpose of the present study is to explore the intrinsic limitations of Rydberg-EIT field sensing in vapor cells, no collision terms or dephasing terms are added. While Rydberg-atom collisions, Penning ionization, and ion electric fields can cause dephasing, such effects can be alleviated by reducing the beam intensities, lowering the vapor pressure, or limiting the atom-field interaction time. In this analysis we set,
$\Gamma_1=0$, $\Gamma_2=2\pi\times$4.56~MHz,
$\Gamma_3=2\pi\times$1.8~MHz, $\Gamma_{4,5}=2\pi\times$(1.3,~2.6)~kHz.
Note, $\Gamma_{2}$ is for the D1 line in $^{133}$Cs, and $\Gamma_{4}$, $\Gamma_{5}$, are typical Rydberg decay rates \cite{SteckCsData, vsibalic2017arc}.

We numerically solve these equations to find the steady-state solution for $\rho_{21}$ for various input values of $\Omega_{P,D,C,RF}$. This is done by forming a matrix with the system of equations for $\dot{\rho}_{ij}=0$. The null-space of the resulting system matrix is the steady-state solution.  The steady-state solution for $\rho_{21}$ is then Doppler averaged\cite{berman2011book}
\begin{equation}
\rho_{21_D}=\frac{1}{\sqrt{\pi}\,\, u}\int_{-3u}^{3u}\rho_{21}\left(\Delta'_P,\Delta'_D\right)\,\,e^{\frac{-v^2}{u^2}}\,\,dv\,\,\, ,
\label{doppler}
\end{equation}
where $u=\sqrt{2k_B T/m}$ is the velocity and $m$ is the mass of the atom. We can achieve convergence to the solution within three standard deviations of the mean velocity. We use the case where the probe and coupling laser are co-propagating. Thus, the frequency seen by the atom moving toward the probe beam is upshifted by $2\pi v/\lambda_P$ (where $v$ is the velocity of the atoms), the frequency of the dressing beam seen by the same atom is downshifted by $2\pi v/\lambda_D$, and the frequency of the coupling beam seen by the same atom is upshifted by $2\pi v/\lambda_D$  The probe, dressing, and coupling beam detunings are modified by the following:
\begin{equation}
\begin{array}{rcl}
\Delta'_P&=&\Delta_P+\frac{2\pi}{\lambda_P}v\\
\Delta'_D&=&\Delta_D-\frac{2\pi}{\lambda_D}v\\
\Delta'_C&=&\Delta_C+\frac{2\pi}{\lambda_C}v \,\,\, ,
\end{array}
\label{doppler2}
\end{equation}
where the ``sign'' in the second terms on the right-hand-side accounts for ``co-propagating'' or ``counter-propagating'' beams with respect to the probe.

\newpage
\section*{Acknowledgements}
\vspace{-5mm}
\noindent This work was partially funded by the DARPA SAVaNT program and the NIST-on-a-Chip (NOAC) Program. We thank Dr. P. Weichman with BAE for technical discussions.

\vspace{-5mm}
\subsection*{Conflict of Interest}
\vspace{-5mm}\noindent The authors have no conflicts of interests to disclose.

\vspace{-5mm}
\subsection*{Data Availability Statement}
\vspace{-5mm}\noindent The data relevant to the findings of this research project are available from the corresponding author upon reasonable request.

\bibliography{atom_probe_bib}
\end{document}